\documentclass[conference]{IEEEtran}
\IEEEoverridecommandlockouts
\usepackage{cite}
\usepackage{amsmath,amssymb,amsfonts}
\usepackage{algorithmic}
\usepackage{graphicx}
\usepackage{textcomp}
\usepackage{xcolor}
\usepackage{hyperref}
\usepackage{tabularx}
\usepackage{booktabs}
\usepackage{float}

\def\BibTeX{{\rm B\kern-.05em{\sc i\kern-.025em b}\kern-.08em
    T\kern-.1667em\lower.7ex\hbox{E}\kern-.125emX}}
\begin{document}


\title{Analyzing Skill Element in Online Fantasy Cricket}



\author{\IEEEauthorblockN{Sarthak Sarkar\IEEEauthorrefmark{1}, Supratim Das\IEEEauthorrefmark{1}, Purushottam Saha\IEEEauthorrefmark{1}, Diganta Mukherjee\IEEEauthorrefmark{2} and Tridib Mukherjee\IEEEauthorrefmark{3}}
\IEEEauthorblockA{\IEEEauthorrefmark{1}Indian Statistical Institute, Kolkata}
\IEEEauthorblockA{\IEEEauthorrefmark{2}Sampling and Official Statistics Unit(SOSU)\\
Indian Statistical Institute, Kolkata}
\IEEEauthorblockA{\IEEEauthorrefmark{3}Chief Data Scientist and AI Officer, IDfy}}

\maketitle

\begin{abstract}

Online fantasy cricket has emerged as large-scale competitive systems in which participants construct virtual teams and compete based on real-world player performances. This massive growth has been accompanied by important questions about whether outcomes are primarily driven by skill or chance. We develop a statistical framework to assess the role of skill in determining success on these platforms. We construct and analyze a range of deterministic and stochastic team selection strategies, based on recent form, historical statistics, statistical optimization, and multi-criteria decision making. Strategy performance is evaluated based on points, ranks, and payoff under two contest structures Mega and 4x or Nothing. An extensive comparison between different strategies is made to find an optimal set of strategies. To capture adaptive behavior, we further introduce a dynamic tournament model in which agent populations evolve through a softmax reweighting mechanism proportional to positive payoff realizations. We demonstrate our work by running extensive numerical experiments on the IPL 2024 dataset. The results provide quantitative evidence in favor of the skill element present in online fantasy cricket platforms.

\end{abstract}


\section{Introduction}
Cricket, one of the most celebrated sports in the world, has a unique cultural and emotional significance for millions of fans. Played between two teams of eleven players each, the game encompasses the core elements of batting, bowling, and fielding, with the central objective of outscoring the opponent through strategic play and athletic performance. In recent years, the traditional experience of following cricket has been revolutionized by the emergence and rapid expansion of fantasy sports platforms such as \textit{Dream11}, \textit{My11Circle}, and others. These platforms enable users to create virtual teams consisting of real-life players and compete against others based on the actual real-time performance of these players on the field. Participants accumulate points according to player match statistics, such as runs scored, wickets taken, or catches held, and compete for rankings and attractive prizes that could be in cash, upgrades (freemium versions) etc.


What initially began as a casual and engaging way to enhance the viewing experience has evolved into a multibillion-dollar industry, reshaping how fans interact with the sport. The accessibility of mobile technology, widespread internet penetration, and the growing passion for cricket in countries such as India have fueled this unprecedented boom. Today, millions of users actively participate in fantasy contests before every major match or tournament, often employing statistical analysis and strategic reasoning to optimize their team selections. However, this rapid rise of fantasy cricket has also led to complex legal, regulatory, and ethical debates. Central to these discussions is the contentious question of whether fantasy sports should be categorized as a form of gambling, driven primarily by chance, or as a game of skill, grounded in knowledge, strategy, and decision-making. This distinction has profound implications for the industry’s legality, taxation, and social perception, particularly in jurisdictions where gambling is heavily regulated or prohibited. As fantasy sports continue to blur the boundaries between entertainment, competition, and commerce, understanding their impact and classification has become increasingly vital for policymakers, economists, and sports analysts alike.

In recent years, online fantasy gaming platforms have attracted significant academic and industry attention, leading to a growing body of research focused on optimizing player selection and improving team performance prediction models. Despite this surge of interest, much of the existing literature has focused mainly on developing algorithms and analytical frameworks to identify the \textit{best eleven} players most likely to perform well on these platforms, while comparatively less emphasis has been placed on exploring the underlying skill component that differentiates the need for skill in such fantasy gaming from pure chance-based gambling type of games. Neither has there been a comprehensive study of the comparison between the existing methods of selecting a team to identify an optimal approach which will give a participant a consistent edge over others. The process of forming a fantasy cricket team involves navigating various constraints, which makes it a complex decision-making problem. Several studies have proposed innovative computational approaches to address this challenge (see the review section \ref{review}). We explicitly add a strategic consideration layer to this analysis.


In this paper, our objective is to investigate the \textit{skill element} present in fantasy cricket contests. We examine whether strategic decision making, analytical reasoning, and informed player selection can consistently outperform purely chance driven outcomes. We will also draw comparisons between a wide range of methods to identify an optimal set of strategies that will perform well in various scenarios. To achieve this, we design a series of experiments that simulate real-world fantasy contests and evaluate how different team formation strategies perform relative to a random selection benchmark. Selecting a team of 11 players at random requires no skill and is based purely on chance, and is therefore selected as our benchmark strategy. To measure and compare performance, we develop a comprehensive simulation framework that replicates the structural and scoring mechanisms of popular online fantasy gaming platforms. The framework utilizes extensive simulation calibrated with real-world match data to create a realistic template. Accurately capturing actual player performances, scoring rules, and contest formats. Within this environment, we construct multiple teams based on different strategic methodologies and then simulate their outcomes in numerous contests. The success of each team is evaluated using a set of quantitative performance metrics that encompass total fantasy points, average rank achieved, and monetary payoff earned under some realistic contest structures.

The experimental setup includes a controlled population of simulated participants, or agents, each representing a distinct strategic behavior. Each agent adheres to a single team selection strategy throughout the duration of the simulated tournament, allowing for a clean and interpretable comparison of outcomes. The analysis is divided into several stages. In the first stage, we evaluate strategies purely on the basis of points scored and rankings obtained, providing an initial measure of their relative performance efficiency. In the second stage, we introduce two distinct contest formats: the Mega Contest and the 4x or Nothing Contest, each with its own payoff structure designed to mimic the prize distribution schemes commonly observed on fantasy sports platforms. The simulation results under these different payoff schemes show how the risk–reward trade-offs of each strategy manifest in terms of realistic contest outcomes. 

We focus on two kinds of payoffs: \textit{player-specific} payoff, which assesses individual agent outcomes, and \textit{strategy-specific} payoff, which assesses the overall effectiveness of a given selection method across multiple participants or contests. Finally, we extend the analysis through a dynamic framework that examines how the distribution of strategies among agents evolves over successive tournaments. This allows us to observe whether certain strategies gain dominance or fade over time, offering insight into the adaptive and evolutionary aspects of skill in fantasy sports participation. 

Before moving on to the analysis, we define certain relevant terms. The players who enter the platform with teams will be called \textit{agents}, the players playing the real-life game will be called \textit{players}, and the sports fantasy platforms where the contests are played will be called \textit{games}.

The remainder of the paper is organized as follows. Section 2 presents an overview of the overall framework, introducing the game, the dataset, and the agents used in the analysis. In Section 3, we describe the methodology adopted for the non-payoff study. Section 4 outlines the payoff structure, the contest formats, and the dynamic tournament setup. Section 5 discusses the results obtained from the numerical experiments. Finally, in Section 6 we conclude by highlighting key findings, limitations, and directions for future research.

\section{A Review of Related Work}
\label{review}

\cite{b1} formulated fantasy team selection as an integer optimization problem in which the decision variables represent whether a given player is selected in the eleven. Their objective is to maximize the expected total fantasy points among the selected players while satisfying platform-specific constraints such as the total budget (credits), the number of players in each role (batsman, bowler, allrounder, wicketkeeper), and team composition constraints. They further draw on ideas from portfolio theory (Markowitz optimization) to penalize inconsistent performers and thereby incorporate a risk dimension (i.e., variability in player performance) alongside the expected return. Their findings illustrate how mathematical programming offers a structured alternative to simple methods of team selection, enabling systematic exploration of the feasible selection space and providing decision-makers with an optimal (or near-optimal) set of eleven players under the given constraints.

Building on this foundation, \cite{b2} introduced a multi-criteria decision-making framework(MCDM), employing the TOPSIS method(Technique for Order of Preference by Similarity to Ideal Solution). In their study, a set of performance metrics, including batting averages, strike rates, bowling averages, economy rates, and all-rounder contributions are first evaluated and weighted (via an AHP, Shannon, or Synthesis based approach) and then aggregated to compute a composite score for each player. Using this ranking, the top eleven players for a T20 match are selected. This methodology highlights how a structured weighting of performance indicators, rather than a single expected point measure, can guide rational team formation by explicitly modeling multiple dimensions of player performance (e.g. batting vs bowling vs fielding) and trade-offs among them.

More recently, \cite{b3} applied machine learning techniques, specifically the k-nearest-neighbor (k-NN) algorithm, to the fantasy cricket domain. Their research incorporates contextual features beyond player performance statistics, such as venue‐specific averages, head-to-head records between teams, recent form, and inning-specific performance filters. They use these as input to the k-NN model to predict which players are likely to perform well in a given match. The predicted top players are then combined into candidate eleven-player teams under platform constraints. \cite{b4} and \cite{b5} both focus on applying machine learning techniques to optimize fantasy-cricket team selection for the Indian Premier League (IPL) on Dream11. They use ensemble models such as Random Forest and XGBoost to forecast player fantasy points and recommend high-performing team combinations under platform-specific constraints. Together, these studies highlight a shift towards the use of machine learning methods for selection of fantasy teams that incorporate match-specific context rather than rely solely on historical averages.

\cite{b6} proposed a deep reinforcement learning (DRL) framework to optimize fantasy sports team selection, treating the task as a sequential decision-making problem under constraints such as budget and player roles. Using historical cricket data, the authors modelled the selection process as a Markov Decision Process, where Deep Q-Network (DQN) and Proximal Policy Optimization (PPO) agents learned policies to construct high-scoring fantasy teams. Their results demonstrated that DRL-based methods outperform conventional heuristic and rule-based selection strategies, highlighting the potential of reinforcement learning for adaptive, data-driven team formation in fantasy sports.

Beyond fantasy platforms centered on cricket, \cite{b7} made a significant contribution to the broader understanding of Daily Fantasy Sports (DFS) dynamics through their comprehensive analysis of contests held during the 2017 National Football League (NFL) season. Their research went beyond conventional team selection optimization by incorporating the behavioral and strategic dimensions of player competition within these online ecosystems. Specifically, they developed a probabilistic framework that models opponent team selection strategies using a Dirichlet-multinomial data-generating process, capturing the inherent uncertainty and interdependence of player choices in large DFS contests. In addition, they proposed an optimization framework to determine the optimal team composition for a rational and risk-neutral decision-maker. They introduced a mean–variance optimization algorithm, a technique traditionally used in financial portfolio theory, to balance expected returns (i.e., fantasy points) against risk (i.e., variability in outcomes). By formulating the problem as a series of binary quadratic programs, they demonstrated a mathematically rigorous way to identify the most efficient fantasy team combinations under realistic constraints such as salary caps, positional requirements, and player correlations. Furthermore, their study made a pioneering attempt to quantify unethical practices within fantasy sports ecosystems. By analyzing contest data and participant behavior, they provided empirical estimates of insider trading and collusion, revealing systemic vulnerabilities in the governance of DFS platforms. These findings shed light on the integrity risks associated with fantasy sports markets and underscored the importance of transparency and fair play in maintaining user trust.

Shifting our focus from finding the optimal team to broader research on fantasy sports, \cite{b8} presents a comprehensive bibliometric analysis of fantasy sports research, examining publication trends, citation patterns, and thematic developments across major databases such as Scopus and Web of Science. The study highlights steady growth in scholarly output over the past decade, identifying key contributors, influential journals, and dominant research clusters focused on player motivation, media engagement, behavioral aspects, and the commercialization of fantasy platforms. Wilkins notes that while the field is expanding, it remains relatively underexplored compared to broader sports analytics, with limited cross-cultural, longitudinal, and data-driven studies.

\cite{b9} investigate the motivations and behavioral patterns of online fantasy sports users (FSUs) using Uses and Gratifications theory and Q-methodology. The study identifies five distinct user types which are Casual Players, Skilled Players, Isolationist Thrill-Seekers, Trash-Talkers, and Formatives—based on combinations of engagement, involvement, and motivational profiles. Key motivations driving participation include surveillance (tracking player statistics), arousal (competition and excitement), and skill-oriented engagement, while social interaction and escapism are less influential. The findings provide important insights into user behavior, engagement patterns, and platform design, emphasizing the role of skill and competitive drive in shaping participation in fantasy sports.

\cite{b10} and \cite{b11} investigate the relative roles of skill and chance in daily fantasy sports (DFS) using statistical and computational modeling approaches. \cite{b10} employ stochastic frontier models and variance decomposition on cricket-based DFS data to show that past performance, participant experience, and strategic contest selection significantly drive outcomes, demonstrating that DFS is primarily skill-dominant and offering a decision-support framework for policy and regulatory evaluation. \cite{b11} used variance-based metrics and simulation of random lineups across U.S. fantasy contests to empirically confirm that skilled participants outperform random strategies, establishing the persistence of skill and providing a quantitative skill–luck continuum relevant for legal and regulatory classification. Together, these studies highlight the importance of performance analytics, skill quantification, and statistical modeling in understanding DFS outcomes and informing regulatory and policy-making decisions.

\section{Framework}


\subsection{About The Game}
Dream11 and My11Circle are among the most popular fantasy sports platforms in India, allowing users to create virtual teams and participate in contests based on real-life cricket matches. On these platforms, participants form teams by strategically selecting players from upcoming matches while adhering to specific rules and constraints such as budget limits, player roles, and the maximum number of players allowed from a single team. Each participant selects a total of 11 players and once the real match begins, points are awarded to every team based on the actual on-field performance of the chosen players. The scoring system considers multiple factors, including runs scored, wickets taken, catches held, strike rates, economy rates, and other match-related statistics, ensuring that player contributions are accurately reflected in their fantasy scores. After the match is completed, the teams are ranked according to their total accumulated points and a fixed proportion of the top ranked participants receive rewards according to the contest’s prize structure. The rewards typically vary depending on the size and entry fee of the contest, with higher-stake competitions offering more substantial payoffs to a smaller percentage of participants. Success in these contests is therefore determined by a combination of strategic team selection and accurate performance prediction, while it is also important to note that an element of chance is always present due to factors such as unpredictable performances, weather interruptions, or injuries.

On these platforms, a wide variety of \textit{contests} are available for every cricket match, each designed with distinct structures and reward mechanisms to accommodate different types of participants. These contests vary in terms of the number of participants allowed, the total prize pool, the percentage of agents who receive prizes, the entry fee, and the maximum number of teams that a single participant can register using one account. The diversity of formats allows users to choose contests that align with their risk preferences, budget constraints, and strategic objectives. Despite this variety, the majority of contests offering cash rewards can be broadly categorized into two main types based on their payout structure and reward distribution:
\begin{itemize}
    \item Contests that give cash prizes in decreasing order of rank. Typically, the first prize is the main attraction, often 50–60 times greater than the second prize. The value of subsequent prizes continues to drop steeply with increasing rank, creating a highly top-heavy reward structure. However, to maintain user participation and perceived fairness, the overall prize pool is distributed among a relatively large portion of participants, typically around 50\%–60\% of the total agents. This ensures that many participants receive at least some form of return. Within this group, only the top 1\%–5\% of the players earn substantial profits, while the remaining winners recover their entry fee or gain a small incremental reward. This structure appeals particularly to high-risk, high-reward players who aim for the top spots while accepting that most outcomes may yield minimal or no profit. These are the most common and widely played contests on online fantasy cricket platforms.
    \item Contests that give cash prizes to the top 10\% - 20\% of the participating agents, but the prize is the same for all, regardless of their exact rank within the winning group. Unlike the previous format, there is no steep drop-off in prize value across ranks, providing a more balanced and predictable return profile. Typically, the prize amount in these contests is set around 4–8 times the entry fee, offering participants the opportunity to multiply their investment without requiring a topmost ranking. This format caters to agents who prefer moderate risk exposure and more stable reward probabilities, as the likelihood of finishing within the winning bracket is higher compared to previous contests, although the potential payoff is comparatively lower.  
\end{itemize}
When selecting teams for contests, agents have to keep in mind certain constraints which influence the composition of the selected teams. These constraints are designed to maintain balance and realism, ensuring that each agent’s team reflects the actual dynamics of a real cricket match, rather than an unrestricted selection of top-performing players. The main constraints are summarized as follows: (1) Each team must include at least one batter, one bowler, one wicketkeeper, and one allrounder. This ensures that each team mirrors the structure of an actual squad and prevents teams from being overloaded with players of a single type (all batters or all bowlers). (2) There should be at least one player from each team competing in the cricket match. This prevents a team from being entirely composed of 11 players from the same real-world team.

In addition to these constraint, players are given credit prices and each agent must construct a team where the sum of player credit must be less than 100. However, we found that all the players in each match had a maximum credit price of 9. So, selecting a team of 11 players would naturally take care of this constraint. Hence, we did not consider this constraint in our analysis.

\subsection{Dataset}
Our analysis is centered on IPL 2024. It marked the 17th edition of the tournament and was held from 22 March to 22 May 2024. The competition featured 10 franchise teams, each comprising both domestic and international players, competing in a league-cum-playoff format. The tournament structure involved a round-robin stage, where every team played multiple matches to accumulate points, followed by playoff rounds that included Qualifier 1, Eliminator, Qualifier 2, and the Final to determine the champion. In total, 74 matches were scheduled for the tournament, of which 71 matches were completed successfully, while 3 matches were abandoned without a toss due to adverse weather conditions. The tournament ended with the Kolkata Knight Riders (KKR) emerging victorious after defeating the Sunrisers Hyderabad (SRH) in the final to clinch the trophy. To setup the simulation framework, we obtained data from multiple reliable cricket sources.

Detailed scorecard information for all completed matches of IPL 2024 was obtained through the CricBuzz API available on RapidAPI. This dataset included granular statistics that cover the batting, bowling, and fielding performances of each player in each match. The data captures essential parameters such as runs scored, balls faced, strike rate, wickets taken, overs bowled, economy rate, catches, and run-outs.

Data related to player career statistics for all players in the tournament was obtained from ESPN Cricinfo., one of the most trusted repositories of historical cricket data. This dataset includes long-term performance indicators such as career averages, strike rates, economy rates, and consistency measures in domestic and international formats.  

\subsection{Agents}
As defined earlier, agents refer to participants who enter fantasy sports contests with their selected teams to compete for rewards. The objective of each agent is to build a team that maximizes the expected points or payoff, depending on the structure of the contest and the scoring system. In our analysis, we consider a total of 15 distinct strategies that agents might adopt when forming their teams, each representing a unique approach. The strategies range from simple, rule-based methods to more sophisticated, data-driven techniques. These strategies are designed to capture the diversity of behaviors observed among real-world fantasy sports participants, ranging from casual users relying on intuition to experienced players who leverage statistical models and historical data. 

Fantasy teams are selected only from the playing eleven members, along with the participating impact player of each team. Players who are part of the squad but not included in the matchday lineup, as well as impact players who do not take part in the game, are excluded from the pool of eligible selections. The list of playing members for each team becomes available approximately half an hour before the match begins. Therefore, instead of selecting from the entire squad beforehand, it is advisable for agents to wait until this time to finalize their teams, as they can then identify which players are guaranteed to participate in the match. This approach increases the likelihood of achieving a higher potential score.
The following three characteristics are pivotal in defining any strategy, and we have broadly categorized our strategies based on these, which are defined as follows:\\
\textbf{Variable: }Variable strategies randomize player selection, which means that each time an agent applies this strategy, a different combination of players is generated for the same match. This approach generates a wide range of teams, from exceptionally strong teams to teams that perform poorly. The key advantage of variable strategies lies in their ability to obtain a top-ranked team among the diverse set of teams created, making these strategies advantageous for contests that reward the highest ranks. However, the drawback is that the majority of the teams generated will not be optimal and therefore yield low or zero rewards in most matches. Although it is possible to win huge rewards for an agent using these strategies, the likelihood is very low. Hence, these strategies are inherently \textit{risk-taking}, emphasizing potential high rewards at the expense of consistency and stability. They mimic the behavior of players who are willing to trade off frequent small losses for the possibility of a big win.\\
\textbf{Deterministic:} Deterministic strategies, on the other hand, follow a fixed decision rule that produces the same team composition every time they are applied for a given match. Thus, these strategies eliminate inherent randomness, and agents employing a deterministic strategy will generate identical teams for a particular match. Since only a limited number of teams can be formed using these strategies, the likelihood of obtaining the highest-ranked team is relatively low. However, if such a strategy happens to produce a strong team, it performs well in the second type of contest defined earlier, where the goal is not to achieve the top rank but to finish within a certain proportion of the best-performing agents. Consequently, these strategies can be viewed as risk-averse, prioritizing reliability over volatility. Agents employing deterministic strategies are less likely to experience extreme fluctuations in their outcomes, making these strategies particularly suitable for participants who prefer consistent, low-risk participation over speculative approaches.\\
\textbf{Learning:} In tournaments and series, player performances often vary significantly, some players perform exceptionally well, while others underperform. Therefore, agents should consider the recent performances of players when selecting the fantasy team. Learning indicates strategies that take into account the performance of players in tournaments and series, learn from them, and then pick the team. For example, players who have consistently performed well in recent matches, are given higher selection probabilities in subsequent games. This continuous adaptation allows learning strategies to capture evolving trends such as form fluctuations, emerging talents and learn from experience and improve predictive team selection over time.

Next, we define the 15 strategies that we will be using in the analysis. These strategies range from simple, intuitive, and easy approaches to more sophisticated, data-driven, and method-based approaches. We have tried to diversify the strategy set to capture the various decision-making styles and behaviors that agents may exhibit while selecting their fantasy teams. We aim to represent both casual participants who rely on intuition or luck and skilled participants who apply analytical or algorithmic reasoning. Each strategy adheres to the constraints offered by the fantasy platforms. We also categorize each strategy according to the characteristics defined above. We summarize the strategies as follows :
\begin{itemize}
    \item \textbf{Strategy 1 (Random 1) :} 11 players are selected randomly from the pool of players keeping in mind the constraints. \textbf{(Variable)}
    \item \textbf{Strategy 2 (Fav\_Team) :} The agent picks one of the two teams which he thinks is more likely to win the match. He takes 10 players from the team and the remaining player from the other team. \textbf{(Variable)}
    \item \textbf{Strategy 3 (Allrounder\_Select\_All) :} Allrounders are given preference. The agent picks as many allrounders as he can keeping in mind the constraints. The idea is allrounders are likely to score more as they can both bowl and bat. \textbf{(Variable)}
    \item \textbf{Strategy 4 (MA5) :} We define the form of a player as the average of the points scored by that player over the previous 5 matches. We calculate the form of each player in the player pool and choose the top 11 player according to form and keeping in my the constraints. \textbf{(Deterministic, Learning)}
    \item \textbf{Strategy 5 (Career\_Averages) :} We look at the career batting average and career wickets taken by each player. Then we choose a batsman, a keeper and an allrounder with the highest career batting average and choose a bowler and an allrounder with the highest career wicket. We randomize the rest of the player chosen keeping in mind the constraints. \textbf{(Variable)}
    \item \textbf{Strategy 6 (Tournament\_Stats) :} We choose the player with the highest run, highest wickets and highest boundaries in the tournament among the pool of players and randomize the rest keeping in mind the constraints. \textbf{(Variable, Learning)}
    \item \textbf{Strategy 7 (Career\_Points) :} We define \textbf{career point} for each player as follows : we take career stats such as runs, fours, sixes, wickets, maidens, catches, stumps and calculate the points scored by these stats based on my11circle point system and divide them by 'innings batted'
    for batting stats, 'innings bowled' for bowling stats and 'innings fielded'(taken as the max of 'innings batted' and 'innings bowled') for fielding stats. Then we rank the players based on the \textbf{career point} and take the best 11 keeping in mind the constraints. \textbf{(Deterministic)}
    \item \textbf{Strategy 8 (Random 2) :} 11 players are selected randomly from the pool of players with a change in the constraints as the number of batsmen selected is minimum 2.
    \textbf{(Variable)}
    \item \textbf{Strategy 9 (MA1) :} We define the form of a player as the points scored by that player in the previous  match. We
calculate the form of each player in the player pool and choose the top 11
player according to form and keeping in my the constraints. \textbf{(Deterministic, Learning)}
    \item \textbf{Strategy 10 (Allrounder\_Pref) :} The players are selected similar to MA1 with an improved constraint as the minimum number of allrounders selected is 3. \textbf{(Deterministic, Learning)}
    \item \textbf{Strategy 11 (Mean\_Var\_Optimization) :} The players are selected using form in last 3 matches as an estimate for performance and penalize
    it using the variance of the form. The degree of the penalization is decided by the risk averse factor. The final team is selected using integer programming optimization problem keeping in mind the constraints. \textbf{(Deterministic, Learning)}
    \item \textbf{Strategy 12, 13, 14 (TOPSIS\_Synthesis, TOPSIS\_AHP, TOPSIS\_Shannon)} : Topsis method gives an ordering to the players based on some performance factors for batsmen, bowlers and allrounders. The order is selected based on multiple criteria and giving weights to the performance factors. The weights are given by AHP, Shannon entropy and Synthesis methods. After this the top ranked players are selected keeping in mind the constraints. \textbf{(Deterministic)}
    \item \textbf{Strategy 15 (Popularity\_Selection):} We define the popularity of a player as the percentage of agents who have selected the player on their team. We choose the top 11 players with the highest popularity at the time of selection, keeping in mind the constraints. \textbf{(Variable)}
\end{itemize}

\section{Methodology}
For our analysis, we sequentially examine each match of the IPL 2024 season to evaluate and compare the performance of different strategic methods. For every match, a fixed number of agents are introduced into the simulated contest environment. Each agent represents a unique participant on a fantasy sports platform and selects their team according to a specific strategy. Once the teams are formed, the simulation is executed using the actual match outcomes from IPL 2024, and the points and ranks of all agents are computed based on the same scoring system used by leading platforms such as Dream11 and My11Circle. We conduct a small-scale analysis, where we introduce 100 agents for each strategy in every match. These agents compete directly against each other under identical conditions. While deterministic strategies produce the same team composition across all 100 agents, maintaining an equal number of participants per strategy ensures uniform sampling and eliminates potential biases arising from unequal representation. We do not consider Popularity\_Selection in this part of the analysis and it will be considered later.

Once all match simulations are completed, we move to the evaluation phase, where our primary objective is to identify and characterize high-performing strategies. To do so, we define several quantitative performance indicators that enable us to distinguish strong strategies from weaker ones. Two key indicators are considered: (i) the number of matches in which an agent using that strategy achieves the top rank, and (ii) the number of matches in which agents using that strategy have the best average rank. These measures help us assess both peak performance and consistency. To obtain deeper insights, we focus on two fundamental aspects of contest outcomes: points and ranks. Since higher points and lower ranks are directly associated with greater rewards on fantasy sports platforms, they serve as natural and interpretable metrics for comparison. For each match, we compute the average rank(average of the ranks obtained by the agents of a strategy), average points(average point scored by the agents of a strategy), and best rank(lowest rank scored by the agents of the strategy) of every strategy.

\subsection{Metrics}
For our analysis, we focus on eight key performance metrics designed to evaluate and compare the relative effectiveness of the strategies used in the simulation. These metrics collectively capture both head-to-head stats of the strategies and also the central moments of key indicators such as ranks and points accumulated by agents using each strategy. By analyzing these dimensions together, we obtain a balanced assessment that considers not only how often a strategy performs best, but also how consistently it performs across matches. The eight metrics are defined as follows:
\begin{itemize}
    \item \textbf{Win\%(Best Rank) :} Percentage of matches the strategy has been ranked 1. A higher value indicates the strategy has a high capability of producing the best-performing team. 
    \item \textbf{Win\%(Average Rank) :} Percentage of matches the average rank of the agents of the strategy was the lowest. A higher value indicates a sense of overall dominance and consistency.
    \item \textbf{Mean(Average Points) :} For each match, average points for a strategy is defined as the mean of the points scored by the agents following the strategy. This metric is the mean of these values across all 71 completed IPL 2024 matches.
    \item \textbf{50\%(Average Points) :} This metric is the median of the average points calculated, across all 71 matches.
    \item \textbf{Mean(Average Rank) :} For each match average rank for a strategy is defined as the mean of the ranks obtained by the agents following the strategy. This metric is the mean of the average ranks over 71 matches.
    \item \textbf{50\%(Average Rank) :} This metric is the median of the average ranks over 71 matches.
    \item \textbf{Mean(Best Rank) :} For each match, best rank for a strategy is defined as the highest rank obtained by the agents following the strategy. This metric is the mean of the best ranks over 71 matches.
    \item \textbf{50\%(Best Rank) :} This metric is the median of the best ranks over 71 matches.
\end{itemize}

The metrics defined above can have different range of values. Hence, to maintain uniformity,we \textit{transform} the data so as to make all the values between 0 and 1. We have as our data matrix a 14 $\times$ 8 table where the rows signify the strategy and the columns signify the metrics. For each metric (column), we first calculate its minimum and maximum values across all strategies. Each entry $x_{ij}$ in the column is then transformed as 
\begin{equation}
x'_{ij} = \frac{x_{ij} - (\min_i x_{ij} - 1)}{\max_i x_{ij} - \min_i x_{ij} + 2} \label{eq}
\end{equation}
This ensures that all values lie in (0,1). For metrics based on ranks (where a lower value is better), we apply an additional transformation 
\begin{equation}
x''_{ij} = 1 - x'_{ij} \label{eq}
\end{equation}
so that higher values consistently represent better performance across all metrics.

Once we have defined the metrics, the next step is to derive a composite scoring system that can effectively summarize and compare the performance of the different strategies in a single, interpretable measure, helping us rank the strategies and identify the best-performing ones. While each of the eight metrics provides valuable insights into specific aspects of performance, there exists a high degree of correlation among them. For instance, strategies that achieve higher average points tend to also secure better ranks and higher win percentages.

To address this, we construct a simple yet meaningful summary measure called the \textit{Average 4}. The score is calculated by taking the simple average of 4 out of the 8 metrics — Win\%(Best Rank), Win\%(Average Rank), Mean(Average Points), and Mean(Best Rank). Each of the four selected metrics has been normalized to lie within the range (0,1) through the transformation procedure discussed earlier. Consequently, the resulting \textit{Average 4} score for each strategy also lies within the same range.

A higher score (closer to 1) indicates that the strategy performs consistently well across multiple key dimensions, demonstrating a combination of frequent top finishes, strong average scoring ability, and reliable rank outcomes. Conversely, a lower score (closer to 0) implies that the strategy performs weakly across most performance dimensions. Hence, \textit{Average 4} offers a simple yet robust comparative tool that effectively captures the multi-dimensional nature of success in fantasy sports strategy evaluation. This composite score successfully distinguishes strong, skill-based strategies from weaker, chance-driven ones.

\subsection{Stratification Based on PCA}

Supratim will write in a concise manner. Or may remove it if optimal strategy finding is not considered.

\subsection{Subset Competition}
The rank metric \textit{Average 4} establishes an overall ranking of the strategies based on the scores. Another approach to assess the relative effectiveness of strategies is to divide them into smaller, homogeneous  groups and allow them to compete within their respective groups. Within each subset, we analyze which strategies consistently achieve the highest values for \textit{Win\%(Best Rank)} and \textit{Win\%(Average Rank)}. We define a strategy as \textit{uniformly better} than another if it outperforms the other in both of these metrics. In other words, a uniformly better strategy demonstrates superior consistency in achieving top ranks as well as in maintaining a strong overall average performance. Such a strategy is thus indicative of robust and reliable performance across different contests, reflecting a higher degree of skill and stability. We make the subsets such that the strategies in each subset are similar to each other. The rationale is that if in each subset a strategy clearly outperforms another similar strategy, then we can disregard the worse strategy and focus only on the \textit{uniformly better} strategy class. The subsets are defined as follows:
\begin{itemize} 
    \item Subset 1 - Random 1, Random 2, Fav\_Team 
    \item Subset 2 - Career\_averages, Allrounder\_Select\_All, Tournament\_stats \item Subset 3 - MA1, MA5, Mean\_var\_optimization, Allrounder\_pref  
    \item Subset 4 - Topsis\_Synthesis, Topsis\_Shannon, Topsis\_AHP, Career\_points 
\end{itemize}

Subset 1 consists of variable strategies that rely primarily on chance, without incorporating any statistical reasoning or historical performance data. Subset 2 includes variable strategies that construct teams using both player career statistics and recent performance metrics. Subset 3 comprises deterministic strategies that are mainly based on the recent form. Subset 4 contains deterministic strategies developed  using custom performance metrics derived from player career statistics.
\section{The Money Game}
Up to this point, our analysis has primarily focused on evaluating strategies through points and ranks achieved by agents. While these indicators provide valuable insights into performance consistency and competitive strength, they do not directly capture the main incentive that drive real-world participation in fantasy sports. The most crucial part of the game is the prize money received by the agents. Every contest on fantasy platforms requires agents to pay an entry fee, which contributes to a common prize pool distributed among top-performing teams based on their final rankings. The distribution of these prizes and the structure of payouts significantly influence player behavior and strategic decision-making. For instance, in contests where only a few top ranks receive large rewards, agents may adopt riskier strategies with higher potential upside, whereas contests with flatter payout structures may encourage safer, more consistent approaches. Hence, we introduce the concept of a \textit{payoff structure}, which explicitly defines both the entry fee and the prize distribution for each contest type. In this analysis, we have defined two broad categories of contests, and accordingly, we will define two different payoff structures for each of them. 
\subsection{Payoff Structure}

We have designed the payoff structure of the contests to keep them as similar to a real-life contest as possible while maintaining practical feasibility in our simulation framework. Due to the relatively small number of participants in our simulations compared to a real-life contest, the entry fee was increased to generate a substantial prize pool and facilitate meaningful analysis to differentiate between strategies in terms of monetary outcomes.
\begin{itemize}
    \item \textbf{Mega Contest :} The entry fee for each agent is 500. There are in total 1500 agents(100 agents from each strategy). The prize distribution is presented in Table 1. The first prize is 50000. The prizes decrease as the ranks increase. The total entry fee collected from all agents combined is 7.5 lakh, and the prize pool is almost 5.3 lakh. So, the prize pool is about 70\% of the total money obtained from the agents. This ensures that a considerable amount of money is distributed as part of cash rewards, and the rest is deducted as platform fee. Cash rewards are given to the top 60\% of the participants, with the majority of lower-ranked winners recovering only their entry fee.  
    \begin{table}[H]
    \centering
    \caption{Prize Structure for Mega Contest}
    \begin{tabular}{@{}ll@{}}
\toprule
\textbf{Rank Range} & \textbf{Prize} \\ \midrule
1st                & 50,000 \\
2nd                & 10,000 \\
3rd                & 5,000 \\
4th                & 2,000 \\
5th                & 1,000 \\
6th -- 10th        & 800 \\
11th -- 25th       & 625 \\
26th -- 50th       & 575 \\
51st -- 100th      & 540 \\
101st -- 300th     & 515 \\
301st -- 599th     & 505 \\
600th -- 900th     & 500 \\
901st -- 1500th    & 0 \\
\bottomrule
\end{tabular}
\end{table}
    \item \textbf{4x or Nothing :} The entry fee for each agent is 100. There are in total 1500 agents(100 agents from each strategy). The prize distribution is presented in Table 2. The prize is the same for all the ranks in the prize range. It is 4 times the entry fee of each agent. The total entry fee for all agents combined is 1.5 lakhs and the prize pool is 1.2 lakhs. So, the prize pool is about 80\% of the total money obtained from the agents. Cash rewards are given to the top 20\% of the participants.

\begin{table}[H]
    \centering
    \caption{Prize Structure for 4x Contest}
\begin{tabular}{@{}ll@{}}
\toprule
\textbf{Rank Range} & \textbf{Prize} \\ \midrule
1st -- 300th       & 400 \\
301st -- 1500th    & 0 \\
\bottomrule
\end{tabular}
\end{table}
\end{itemize}

\subsection{Metric}
Once the payoff structures for each contest are established, we proceed to simulate both contest types by sequentially introducing 1,500 agents(100 agents of the each strategy) into the contests in a random order. This randomization ensures fairness and prevents any potential bias arising from the sequence of participation. Each simulation is run for the entire tournament duration, allowing us to record the performance outcomes of all agents under both payoff structures. The resulting data provide a comprehensive basis for comparing how different strategies perform when exposed to varying reward mechanisms and competition dynamics.

Following the simulations, we conduct two complementary levels of analysis to interpret the results:\\
\textbf{Player Specific :} This approach focuses on evaluating performance at the individual level. We track the progress of each agent throughout the tournament by monitoring their payoffs across all matches. For every agent, we compute the total payoff accumulated over the course of the tournament. These individual totals are then aggregated at the strategy level to provide a comprehensive summary of how each strategy performs overall. The performance of each strategy is summarized by reporting the maximum, mean, minimum, and quartile values of the agents’ total payoffs within that strategy.\\
\textbf{Strategy Specific :} In this approach, the unit of focus shifts from the individual agent to the strategy as a collective entity. For each match of the tournament, we compute the mean, median, maximum, and minimum payoffs of all agents employing a particular strategy. These match-level results are then aggregated across all 71 matches to obtain the overall mean and median payoffs of each strategy throughout the tournament.

\subsection{Dynamic Tournament}
Throughout our analysis, we have maintained an equal representation of all strategies to ensure a fair and controlled comparison of their performance. This uniform distribution allows us to evaluate each strategy’s strengths and weaknesses in isolation, without the influence of population dynamics. However, this setup is an idealized scenario and does not fully reflect how strategies would naturally evolve in real-world or adaptive systems. In practical scenarios, agents are not bound to a fixed strategy. Instead, they adapt over time, often abandoning strategies that consistently yield poor results and adopting those with better performance. This process of strategy evolution results in poorly performing strategies gradually losing representation, while more effective strategies become more prevalent. To capture this phenomenon, we implement a \textit{dynamic tournament} in our simulations.

The tournament simulation is conducted over 100 iterations, with each iteration comprising 71 matches corresponding to the IPL 2024 season. After each iteration, the number of agents assigned to each strategy is reweighted based on observed performance. To introduce variability and reduce potential bias from the fixed sequence of matches, we generate a bootstrap sample (with replacement) of 71 matches for each iteration. The first iteration, referred to as the burn-in phase, serves as a baseline (or 0th stage) of the simulation. It includes 1,400 agents, with 100 agents assigned to each of the 14 strategies, excluding the Popularity\_selection strategy. This stage establishes initial performance trends and does not include adaptive updates. From the second iteration onward, termed Iteration 1, we introduce the Popularity\_selection strategy, increasing the total number of agents to 1,500. From this point on, the distribution of agents across strategies is no longer fixed; it is updated after each iteration based on the performance of each strategy in the previous iteration.

The reweighting metric is the number of agents in a particular strategy who achieve a positive overall payoff in an iteration. This choice is motivated by the principle that agents with positive payoffs are likely to continue using the same strategy, while those with negative payoffs are likely to switch strategies in search of higher returns. Formally, let $x_i$ denote the number of agents using the $i^\text{th}$ strategy who attain a positive overall payoff. We compute the updated weight for each strategy using a softmax-like function:
\begin{equation}
w_i = \frac{e^{x_i / 25}}{\sum_j e^{x_j / 25}}
\end{equation} The number of agents allocated to the $i^\text{th}$ strategy in the subsequent iteration is then given by:
\[
\text{Agent count for strategy } i = w_i \times (\text{Total number of agents})
\] These counts are rounded to the nearest integer, with minor adjustments to ensure the total number of agents remains fixed at 1,500. To enhance the stability and robustness of the reweighting process, we repeat the tournament simulation and reweighting six times for each iteration and take the average of the resulting weights.

\section{Numerical Experiments}
As discussed earlier, our simulations are conducted across all 71 completed matches of the IPL 2024 season. In each match, 100 agents representing every strategy are introduced into the contest, where each agent independently forms a team based on the rules and decision framework of its respective strategy. After the completion of each match, agents are assigned points, ranks, and corresponding payoffs based on their team’s performance outcomes. We divide our analysis into two parts. The first part contains the analysis based on points and ranks only. The second part contains the analysis based on the payoff. 
\subsection{Analysis 1}

\begin{table}[t!]
\caption{Performance Comparison of Different Strategies}
\centering
\begin{tabular}{lcc}
\hline
\textbf{Strategy Name} & \textbf{Win\% (Best Rank)} & \textbf{Win\% (Average Rank)} \\ 
\hline
Random1 & 23.944 & 0 \\
Tournament\_stats & 19.718 & 1.408 \\
Career\_averages & 18.310 & 5.634 \\
Random2 & 14.085 & 0 \\
Allrounder\_Select\_All & 11.268 & 2.817 \\
Fav\_Team & 7.042 & 0 \\
MA5 & 1.408 & 21.127 \\
TOPSIS\_Shannon & 1.408 & 18.310 \\
TOPSIS\_AHP & 1.408 & 11.268 \\
Mean\_var\_optimization & 1.408 & 8.451 \\
Career\_points & 0 & 14.085 \\
TOPSIS\_Synthesis & 0 & 9.859 \\
MA1 & 0 & 5.634 \\
Allrounder\_pref & 0 & 5.634 \\
\hline
\end{tabular}
\label{tab:strategy_comparison}
\end{table}

The metrics Win\%(Best Rank) and Win\%(Average Points) serve as key indicators of strategy performance, reflecting the proportion of matches (out of 71) in which a strategy either secures the top rank or achieves the highest average points. Table 3 presents these two metrics for all strategies. We observe that the variable strategies are ranked first in approximately 95\% of the 71 games, indicating that they frequently produce the top-performing teams. This suggests that variable strategies occasionally generate optimal team combinations that outperform deterministic ones. However, the average rank and the median rank of the variable strategies are very high. The average points and median points of the variable strategies are also very low. This implies that while a few instances of variable strategies yield exceptional outcomes, the majority of teams they produce are weak and perform poorly. Although variable strategies dominate the top positions on occasion, the likelihood of any team generated through them achieving first rank remains relatively low. In contrast, the deterministic strategies exhibit the highest average ranks in roughly 89\% of the 71 matches, suggesting that most teams formed under deterministic frameworks perform consistently well. These strategies demonstrate low average and median ranks, along with higher average and median points, when compared to variable strategies. The best rank achieved by deterministic strategies is also notably strong. This pattern indicates that, while deterministic strategies may not frequently produce the single top-performing team, they consistently generate high-quality teams that achieve strong and stable performances across contests.

If we examine the results at the strategy level, we observe that among the variable strategies, Random 1, Tournament\_Stats, and Career\_Average exhibit high values for Win\%(Best Rank). On the other hand, within the deterministic group, MA5, Topsis\_Shannon, and Career\_Points achieve relatively high values for Win\%(Average Rank). The strong performance of Random 1 in terms of Win\%(Best Rank) can be attributed to its ability to generate a wide range of teams without any additional constraints, which occasionally results in a winning combination purely by chance. However, since its Win\%(Average Rank) value is 0, it is evident that most teams generated by this strategy perform poorly on average. Conversely, MA5 achieves the highest Win\%(Average Rank), highlighting its capability to consistently produce well-balanced teams that perform strongly on average. Among the variable strategies, Career\_Average records the highest Win\%(Average Rank). After computing the values of Average 4 from the transformed data matrix, we rank all strategies in decreasing order based on their respective scores, as presented in Table 4. Career\_Averages achieves the highest Average 4 score, indicating its strong and consistent performance across multiple evaluation dimensions. This is followed closely by Tournament\_Stats and MA5, both of which demonstrate stable and competitive results throughout the tournament. Notably, these three strategies outperform Random 1, highlighting the advantage of incorporating structured, data-driven approaches in fantasy team formation, leveraging player statistics, recent form, and historical data.
\begin{table}[t!]
\caption{Ranking of Strategies on the Basis of Average 4}
\centering
\begin{tabular}{c}
\hline
\textbf{Strategy Name (Average 4)} \\
\hline
Career\_averages \\
Tournament\_stats \\
MA5 \\
Random1 \\
TOPSIS\_Shannon \\
TOPSIS\_AHP \\
Random2 \\
Allrounder\_Select\_All \\
Career\_points \\
TOPSIS\_Synthesis \\
Fav\_Team \\
Mean\_var\_optimization \\
MA1 \\
Allrounder\_pref \\
\hline
\end{tabular}
\label{tab:avg4_ranking}
\end{table}

For the subset competition, we divided the strategies into several groups such that the strategies within each subset are similar. We conducted simulations for each subset independently using the same dataset as in the previous analysis and evaluated the results based on Win\%(Best Rank) and Win\%(Average Points). For subsets containing variable strategies, each simulation produces different outcomes due to the inherent randomness, so the results were aggregated over six independent runs. For subsets containing deterministic strategies, both metrics yield identical values, and hence, it suffices to analyze only one of them. In Subset 1, Random 1 consistently exhibits higher values for both metrics across nearly all six simulations, making it the uniformly better strategy within this group. In Subset 2, both Career\_Averages and Tournament\_Stats outperform Allrounder\_Select\_All across all runs, thus being uniformly better. However, between Career\_Averages and Tournament\_Stats, no definitive ordering can be established, as each outperforms the other in different runs. In Subset 3, MA5 clearly achieves the highest metric values, establishing it as the dominant strategy within its group. Finally, in Subset 4, Career\_Points and Topsis\_Shannon display closely aligned and superior performance compared to the other two strategies in the subset. 

Overall, this analysis highlights a consistent pattern; while randomness occasionally leads to exceptional results, structured, data-driven, and deterministic approaches tend to perform more reliably and demonstrate stronger average performance across matches.

PCA results to be added by supratim. (if needed)

\subsection{Analysis 2}
We first examine the results of the Mega Contest. In the player-specific analysis, we observe that for several strategies such as MA1, Mean\_Var\_Optimization, Allrounder\_ref, Career\_Points, Topsis\_Synthesis, and Popularity\_Selection, the maximum payoff is negative. This implies that all agents employing these strategies have an overall negative payoff, incurring losses throughout the tournament and reflecting their poor performance under this contest structure. Among all strategies, MA5 stands out by exhibiting consistently high payoffs across all metrics, outperforming most variable strategies. Random 1, in contrast, emerges as a highly risky approach as it achieves the highest positive maximum payoff but also produces strong negative mean and median payoffs, reflecting high variability and volatility in outcome across agents employing this strategy. Among the variable strategies, Career\_Averages and Tournament\_Stats demonstrate relatively safer and more stable performance compared to Random 1, suggesting that incorporating player performance data helps mitigate risk. Among deterministic strategies, Topsis\_AHP also shows consistent performance across many metrics, although it is not as good as MA5.

In the strategy-specific analysis, we observe a clear contrast between variable and deterministic strategies. Variable strategies tend to exhibit a much lower minimum payoff, indicating a higher downside risk compared to deterministic strategies. Conversely, their maximum payoffs are significantly higher, highlighting their potential for exceptionally high returns. While the median payoffs are generally similar across both strategy types, differing only in a few cases, the mean of the average payoff is also relatively close. A key distinction emerges in the median of the average payoff, which is negative for random strategies and positive for deterministic ones. This further suggests that, on average, deterministic strategies offer more consistent and stable performance, while variable strategies carry greater variability and risk.

In the dynamic tournament, the simulation conducted over 100 iterations is illustrated in Figure 1. The figure contains only the top six strategies based on their agent distribution. Over successive iterations, we observe a gradual and systematic redistribution of agents across strategies. The population is largely dominated by variable strategies, with MA5 emerging as the only deterministic strategy maintaining a strong and stable presence throughout. Additionally, the distribution of agents exhibits a periodic pattern, suggesting that the relative attractiveness of strategies fluctuates over time as agents adapt their choices based on payoff outcomes. 

We next turn to the 4x or Nothing contest results. In the player-specific analysis, several strategies including MA1, Mean\_Var\_Optimization, Allrounder\_Pref, Career\_opints, and Popularity\_Selection again show negative maximum payoffs. This implies that all agents employing these strategies have an overall negative payoff, reflecting their poor performance under this contest structure. MA5 performs exceptionally well in this contest structure, as its minimum payoff is positive, meaning that every agent employing MA5 ends the tournament with a profit. Among the variable strategies, Career\_Averages and Tournament\_Stats remain somewhat risky as they exhibit large negative minimum payoffs, but their mean, median, and maximum payoffs remain competitive, even surpassing those of several deterministic strategies. Within the deterministic group, Topsis\_AHP again performs strongly, though it does not quite match the consistency of MA5.

In the strategy-specific analysis, we observe patterns similar to those in the Mega Contest. Random strategies continue to exhibit wider variability, with extreme values on both ends, while deterministic strategies maintain narrower and more stable ranges of payoffs. Notably, MA5 and Topsis\_AHP stand out with high average and median payoffs, reaffirming their status as the most effective and consistent performers. In contrast, most other strategies yield negative mean and median values, indicating that they are less effective within this model.

We ran the dynamic tournament in the 4x or Nothing contest setup multiple times, as the distribution of agents varied across simulations. A consistent pattern emerged: each simulation completed within 3–5 iterations,ultimately converging to a clear majority strategy that dominated the agent population. After the initial burn-in phase, deterministic strategies dominated the early iterations. However, as the iterations progressed, there was a sharp redistribution of agents, and the final population concentrated around only two or three top-performing strategies. In nearly all simulations, a single strategy emerged as the dominant one, most often MA5 or MA1. We show the result of one such simulation in Fig 2. The simulation completed in the second iteration and a clear majority for MA1 was obtained. This shows the strength of using form in selecting a team for these type of contests.

An interesting and somewhat counterintuitive finding arises from the performance of the Popularity\_Selection strategy, which performs poorly in both contest types. Intuitively, one might expect this strategy which selects the most popular players in the contests i.e., the players who have been selected by the most agents for a particular contest, to produce at least average results, since most agents are also selecting these players. However, the poor outcome can be attributed to the equal representation of all strategies in the simulation. Because no filtering mechanism excludes weaker strategies, many poorly performing teams enter the contests, and their selections contribute heavily to overall player popularity. Consequently, the most popular selections often originate from these suboptimal strategies, leading to inferior performance despite the perceived strength of its selections.

\begin{figure}[t!]
\centering
\includegraphics[width=0.95\linewidth]{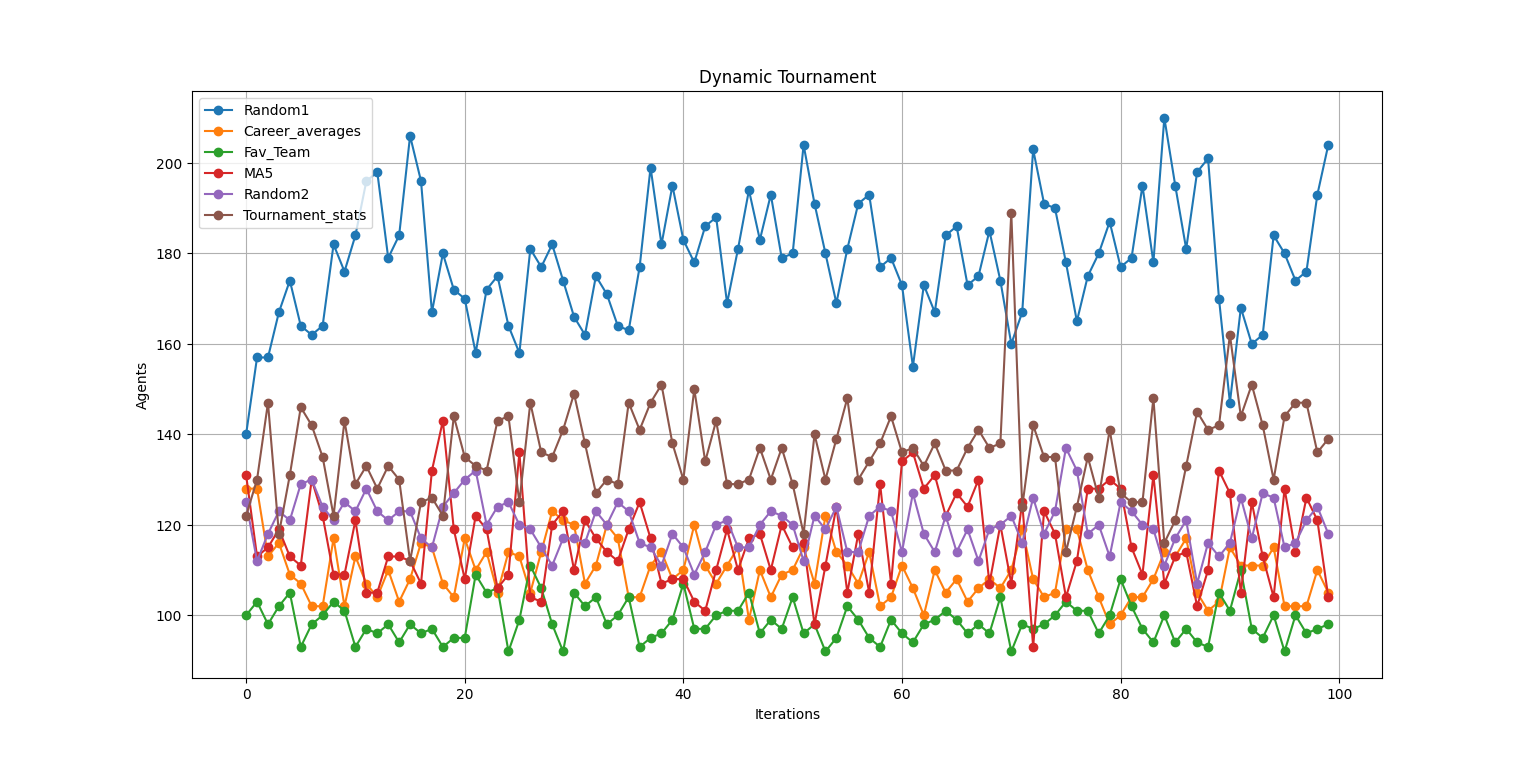}
\caption{Top 6 Strategies of Mega Contest}
\label{fig:plot1}
\end{figure}

\section{Conclusion}

In this paper, we introduced a range of intuitive and easily implementable strategies for selecting a team of eleven players in online fantasy sports platforms. Our primary goal was to demonstrate that even simple, data-driven, and logically constructed strategies can outperform purely random team selections. This finding directly supports the argument that fantasy sports involve a significant element of skill, rather than being games of pure chance. Based on the analysis of points and ranks, our findings indicate that while no single strategy consistently outperformed the Random 1 approach in terms of winning the maximum proportion of matches, strategies such as Tournament\_Stats and Career\_Averages achieved performances that were very close to it. Tournament\_Stats and Career\_Averages also had a higher value than Random 1 across many other metrics, especially Average 4 and thus it can be considered that these two demonstrate better performance than Random 1. Moreover, several other strategies showed superior outcomes to Random 1 across other dimensions, such as average ranking and consistency. In particular, MA5 and Topsis\_Shannon emerged as standout performers, producing teams that, on average, outscored those generated by Random 1. These results highlight that deterministic and statistically informed strategies tend to offer lower risk and higher reliability, providing agents with more stable outcomes across contests. Consequently, for an individual participant, Random 1 represents a high-risk, low-reward strategy with minimal chances of consistent success. On the basis of payoff, we have a more definitive answer. MA5 demonstrated remarkable robustness across both contest types, consistently outperforming Random 1 in multiple performance metrics. Similarly, Tournament\_Stats and Career\_Averages strategies also achieved superior results compared to Random 1, confirming that historical and contextual player performance data can meaningfully enhance team selection quality. Together, these findings provide strong evidence that applying systematic and analytical thinking to fantasy team formation significantly improves the probability of success. Hence, the outcomes substantiate the claim that online fantasy platforms are indeed games of skill, where informed choices and strategic reasoning play a crucial role, as opposed to being governed purely by luck.
\begin{figure}[t!]
    \centering
    \includegraphics[width=0.9\linewidth]{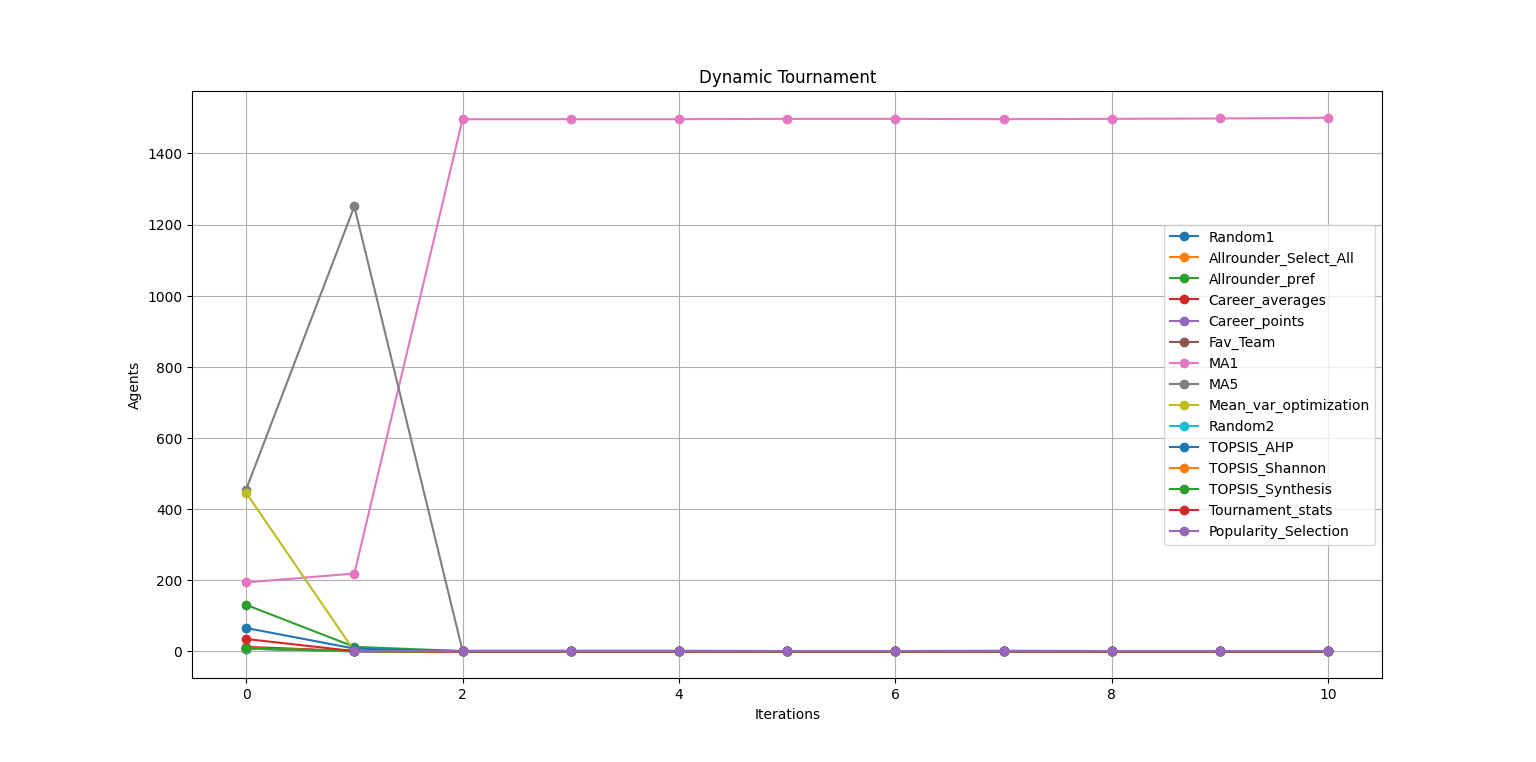}
    \caption{All strategies of 4x or Nothing}
    \label{fig:plot4}
\end{figure}
The main purpose of this paper was to use very simple and intuitive strategies to demonstrate a level of skill on online fantasy platforms and to identify the best-performing strategies by creating an environment similar to an
existing one. Although we tried to replicate the online fantasy platforms as closely as possible, we were unable to do so due to technical constraints. For instance, our analysis did not account for in-match substitutions or scenarios involving multiple players participating in a run-out. The player hitting the wickets in a run-out is given 12 points as opposed to the usual 12 points distributed equally between the player who throws the ball and the player who hits the stumps in a run-out. Additionally, as indicated earlier due to data limitations, only the starting eleven and the impact player from each team were included in the selection pool, rather than the entire squad. Since it is optimal to select from the starting eleven only as argued earlier, this does not make much of a difference. Since the complete list of potential impact players was not available, only those who actually participated in matches were considered. Furthermore, contextual variables such as player credit values, pitch conditions, and average scores that often influence decision-making on real fantasy platforms were not incorporated into the current framework. We also note that the payoff structure adopted in this study represents just one possible formulation among many. Alternative payoff schemes exist across both contest types, and experimenting with these could yield further insights into how scoring systems influence strategic behavior. Additionally, in our simulation setup, if two agents achieved identical scores, the agent entering the contest earlier was assigned the higher rank.

Looking ahead, several promising directions for future research emerge from this work. First, more complex and adaptive strategies could be incorporated to strengthen the empirical evidence supporting the skill-based nature of fantasy sports. Introducing dynamic agents capable of changing their strategies during the tournament based on performance feedback could more accurately model real-world player behavior and provide deeper insights into strategic xevolution. Moreover, extending the framework to other leagues or tournaments beyond IPL 2024, such as international series or franchise-based competitions in different sports, would help test the generalization and robustness of our conclusions. Finally, integrating richer contextual data such as real-time player form, weather conditions, credit points could further bridge the gap between simulation and reality, offering a more comprehensive understanding of skill, strategy, and decision-making in fantasy sports.

\end{document}